\documentclass[conference]{IEEEtran}
\IEEEoverridecommandlockouts
\usepackage{cite}
\usepackage{amsmath,amssymb,amsfonts}
\usepackage{graphicx}
\usepackage{textcomp}
\usepackage{xcolor}

\usepackage{amssymb}
\usepackage{balance}
\usepackage{bm}
\usepackage{adjustbox}
\usepackage{graphicx,xcolor} 
\usepackage[flushleft]{threeparttable}
\usepackage{balance}
\usepackage{algorithm}
\usepackage{algpseudocode}
\usepackage{multirow}
\usepackage{booktabs}
\usepackage{listings}
\usepackage{xcolor}
\usepackage{amsmath}
\usepackage{algpseudocode}
\usepackage{tabularx,booktabs}

\usepackage{listings}

\definecolor{codegreen}{rgb}{0,0.6,0}
\definecolor{codegray}{rgb}{0.5,0.5,0.5}
\definecolor{codepurple}{rgb}{0,0,0}
\definecolor{backcolour}{rgb}{0.95,0.95,0.98}

\lstdefinestyle{mystyle}{
    backgroundcolor=\color{backcolour},   
    commentstyle=\color{codegreen},
    keywordstyle=\color{black},
    numberstyle=\tiny\color{codegray},
    stringstyle=\color{codepurple},
    basicstyle=\ttfamily\footnotesize,
    breakatwhitespace=false,         
    breaklines=true,                 
    captionpos=b,                    
    keepspaces=true,                 
    numbers=left,                    
    numbersep=5pt,                  
    showspaces=false,                
    showstringspaces=false,
    showtabs=false,                  
    tabsize=2
}

\algnewcommand\algorithmicswitch{\textbf{switch}}
\algnewcommand\algorithmiccase{\textbf{case}}
\algnewcommand\algorithmicassert{\texttt{assert}}
\algnewcommand\Assert[1]{\State \algorithmicassert(#1)}%
\algdef{SE}[SWITCH]{Switch}{EndSwitch}[1]{\algorithmicswitch\ #1\ \algorithmicdo}{\algorithmicend\ \algorithmicswitch}%
\algdef{SE}[CASE]{Case}{EndCase}[1]{\algorithmiccase\ #1}{\algorithmicend\ \algorithmiccase}%
\algtext*{EndSwitch}%
\algtext*{EndCase}%

\def\BibTeX{{\rm B\kern-.05em{\sc i\kern-.025em b}\kern-.08em
    T\kern-.1667em\lower.7ex\hbox{E}\kern-.125emX}}
\begin{document}

\title{HyScale-GNN: A \underline{Scal}abl\underline{e} \underline{H}ybrid \underline{GNN} Training System on Single-Node Heterogeneous Architecture}


\author{\IEEEauthorblockN{Yi-Chien Lin,
Viktor Prasanna }
\IEEEauthorblockA{University of Southern California, Los Angeles, California\\
Email: {\{yichienl, prasanna\}}@usc.edu }}

\maketitle

\begin{abstract}
Graph Neural Networks (GNNs) have shown success in many real-world applications that involve graph-structured data.
Most of the existing single-node GNN training systems are capable of training medium-scale graphs with tens of millions of edges;
however, scaling them to large-scale graphs with billions of edges remains challenging.
In addition, it is challenging to map GNN training algorithms onto a computation node as state-of-the-art machines feature heterogeneous architecture consisting of multiple processors and a variety of accelerators.

We propose HyScale-GNN, a novel system to train GNN models on a single-node heterogeneous architecture.
HyScale-GNN performs hybrid training which utilizes both the processors and the accelerators to train a model collaboratively.
Our system design overcomes the memory size limitation of existing works and is optimized for training GNNs on large-scale graphs.
We propose a two-stage data pre-fetching scheme to reduce the communication overhead during GNN training.
To improve task mapping efficiency, we propose a dynamic resource management mechanism, which adjusts the workload assignment and resource allocation during runtime.
We evaluate HyScale-GNN on a CPU-GPU and a CPU-FPGA heterogeneous architecture.
Using several large-scale datasets and two widely-used GNN models, we compare the performance of our design with a multi-GPU baseline implemented in PyTorch-Geometric.
The CPU-GPU design and the CPU-FPGA design achieve up to $2.08\times$ speedup and $12.6\times$ speedup, respectively.
Compared with the state-of-the-art large-scale multi-node GNN training systems such as $P^3$ and DistDGL, our CPU-FPGA design achieves up to $5.27\times$ speedup using a single node.

\end{abstract}


\section{Introduction}\label{sec:intro}
Graph Neural Networks (GNNs) have become state-of-the-art models for representation learning on graphs, facilitating many applications such as molecular property prediction \cite{yang_li_2023,graphsage}, social recommendation system \cite{recommend1,recommend2}, electronic design automation \cite{gnn-eda,eda2}, etc. 
These domains often involve large-scale graphs with over billion edges \cite{hu2021ogblsc}.
Scaling GNN training systems to support such large graphs remains challenging.
Previous works \cite{carla,graphact,h-gcn, linGCN} store the input graph in the device memory (e.g., GPU global memory, FPGA local DDR memory) rather than the CPU memory because accessing data from device memory via DDR channel is much faster than accessing data from the CPU memory via PCIe.
The drawback of this setup is that the size of the device memory is limited (usually 16 to 64 GB), so it cannot accommodate large-scale graphs such as MAG240M \cite{hu2021ogblsc} (202 GB);
storing the graph in the CPU memory can overcome this limitation, 
but then the data needs to be fetched via PCIe which has lower memory bandwidth.
In addition to memory size limitation, it is also challenging to map GNN training algorithms onto a target platform because of the complex architecture of modern machines.
In particular, state-of-the-art nodes adopt a heterogeneous architecture design to meet the performance requirements of various applications \cite{aws,azure}.
A heterogeneous architecture consists of multiple multi-core CPUs connected to several accelerators; the connected accelerators can be GPUs, FPGAs, or AI-specific accelerators \cite{groq,tpu,vpu}.
Most of the existing works adopt a naive and static task mapping \cite{graphact, hp-gnn, pyg} which treats the CPU as a preprocessor, whose main purpose is to offload GNN computations to the accelerator;
this straightforward task mapping overlooks the potential of utilizing the CPU resources for training.
{For example, on a dual-socket AMD EPYC 7763 (7.2 TFLOPS) platform equipped with a single Nvidia RTX A5000 (27.8 TFLOPS), utilizing CPU+GPU for training can potentially provide a (7.2+27.8) / 27.8 = 1.26$\times$ speedup compared with GPU-only training.
In addition, the speed of executing GNN training tasks depends on both the training algorithm and the performance of the target platform;
this makes static task mapping inefficient.}


Motivated by the challenges, we propose HyScale-GNN, a \textit{hybrid} GNN training system that can efficiently train GNN models on a given heterogeneous architecture.
We propose a general processor-accelerator training protocol, which defines how the processors and the accelerators should interact and synchronize to collaboratively train a GNN model.
The protocol is generic and can be adapted to various accelerators including GPU, FPGA, or AI-specific accelerators.
We propose a dynamic resource management mechanism to efficiently map GNN training tasks onto a given heterogeneous architecture.
The mechanism assigns GNN training tasks to both the CPUs and the accelerators, and dynamically adjusts the workload assignment during runtime.
Unlike previous works that result in CPU idling most of the time, our hybrid training system efficiently utilizes both the CPUs and the accelerators to collaboratively train a GNN model.
In addition, HyScale-GNN supports GNN training on large-scale graphs with billions of edges.
To accommodate large-scale graphs, our system stores the input graph in the CPU memory, which can be several terabytes on high-end nodes.
To mitigate the expensive PCIe communication overhead of reading data from the CPU memory, we propose a two-stage feature prefetching scheme to pre-load the required data onto the accelerator. 
While we apply various optimizations in our system, these optimizations do not alter the semantics of the GNN training algorithm;
thus, the convergence rate and model accuracy remain the same as the original sequential algorithm. 

\begin{table}[]
\centering
\caption{Notations of GNN}

\begin{adjustbox}{max width=0.485\textwidth}
\begin{tabular}{cc|cc}
\toprule
 \textbf{{Notation}} & \textbf{{Description}}  & \textbf{{Notation}}  & \textbf{{Description}} \\
 \midrule
\midrule
{$  \mathcal{G}(\mathcal{V},\mathcal{E})$ }& {input graph topology}  & $ \bm{h}_{i}^{l}$& feature vector of $ v_{i}$ at layer $l$\\ \midrule
$ \mathcal{V}$ &  {set of vertices} &     $ \bm{a}_{i}^{l}$& aggregated result of $ v_{i}$ at layer $l$   \\ \midrule
$ \mathcal{E}$& {set of edges} & $ L$ & {number of GNN layers}  \\ \midrule
$\bm{X}$ & input feature matrix & $f^l$ & feature length of layer $l$  \\ \midrule
$ \mathcal{V}^l$& {sampled vertices at layer $l$} &  $ \mathcal{N}(i)$& neighbors of $ v_{i}$ \\ \midrule
$ \mathcal{E}^l$ & {sampled edges at layer $l$}  & $\phi(.)$  & element-wise activation  \\  \midrule
$\bm{X}'$ & feature matrix of sampled vertices & $\bm{W}^{l}$ & weight matrix of layer $l$ \\

\bottomrule
\end{tabular}
\end{adjustbox}
\label{tab:notations}
\end{table}

We summarize the contributions of this work as follows:
\begin{itemize}
    \item We propose HyScale-GNN, a hybrid GNN training system that efficiently utilizes both the CPUs and the accelerators to perform GNN training collaboratively. Our system achieves the same convergence rate and model accuracy as existing works \cite{p3, pagraph, distdgl}  as the proposed optimizations do not alter the original training algorithm.
    \item We propose a general processor-accelerator training protocol that enables HyScale-GNN to work with various accelerators including GPUs, FPGAs, or AI-specific accelerators.
    \item To support GNN training on large-scale graphs (such as ogbn-papers100\cite{ogb} and MAG240M\cite{hu2021ogblsc}), we propose to store the input graph in the CPU memory, and perform two-stage data prefetching to hide the communication overhead.
    \item We propose a performance model which predicts the training performance of our system based on algorithmic parameters of the GNN training algorithm and platform metadata.
    HyScale-GNN utilizes the predicted performance to derive a coarse-grained task mapping onto the target platform during the design phase.
    \item We propose a dynamic resource management mechanism, which performs fine-grained task mapping by fine-tuning the workload assigned to the CPUs and the accelerators during runtime. 
    \item We evaluate HyScale-GNN using several large-scale graphs and two widely used GNN models: GraphSAGE\cite{graphsage}, and GCN\cite{gcn}.
    On a dual-socket platform connected to 4 high-end GPUs, and a dual-socket platform connected to 4 high-end FPGAs, our CPU-GPU and CPU-FPGA designs achieve up to $2.08\times$ speedup, and $12.6\times$ speedup compared with our multi-GPU baseline implemented using PyTorch-Geometric \cite{pyg}, respectively. Compared with the state-of-the-art distributed GNN training systems \cite{p3, distdgl} that use 16 to 64 GPUs on a multi-node cluster, our CPU-FPGA design achieves up to $5.2\times$ speedup using only 4 FPGAs on a single-node.
\end{itemize}


\section{Background}


\subsection{Graph Neural Networks}\label{sec:GNN}
We defined the notations related to a GNN in Table \ref{tab:notations}. 
A GNN learns to generate low-dimensional vector representation (i.e., node embeddings) for a set of vertices (i.e., target vertices $\mathcal{V}^L$), and the node embeddings can facilitate many downstream applications as mentioned in Section \ref{sec:intro}.
A GNN model can be expressed using the aggregate-update paradigm \cite{mp-gnn}:

\begin{equation}\label{eq:agg}
    \bm{a}_v^l = {\text{AGGREGATE}}(\bm{h}_u^{l-1}:u\in \mathcal{N}(v)\cup  \{v\})
\end{equation}
\begin{equation}\label{eq:up}
    \bm{h}_v^l = \phi(\text{UPDATE}(\bm{a}_v^l, \bm{W}^l))
\end{equation}
During feature aggregation, for each vertex $v$, the feature vectors $h_u^{l-1}$ of the neighbor vertices $u \in \mathcal{N}(v)$ are aggregated into $a_v^l$ using algorithm-specific operators such as mean, max, or LSTM.
Since graph-structured data are non-Euclidean, accessing the feature vectors $h_u^{l-1}$ of the neighbor vertices incurs a massive volume of irregular data access.
The feature update phase is a multi-layer perceptron (MLP) followed by an element-wise activation function $\phi$ (e.g., ReLU), which applies a linear transformation and a non-linear transformation to $a_v^l$, respectively. 
While there exist a variety of GNN models, these models follow the  aggregate-update paradigm.
We list two representative models as an example:
\begin{itemize}
    \item GCN \cite{gcn}: is one of the most widely-used GNN models. The model can be specified as follows:
    \begin{equation}
    \begin{split}
        \bm{a}_{v}^{l} & = \text{Sum}(\frac{1}{ \sqrt{D(v)\cdot D(u)}} \cdot \bm{h}_{u}^{l-1} )\\
        \bm{h}_{v}^{l} & = \text{ReLU} \left(\bm{a}_{v}^{l}\bm{W}^{l} + \bm{b}^{l} \right)
    \end{split}
\end{equation}
Where $D(v)$ denotes the degree of vertex $v$, and $\bm{b}^{l}$ indicates the bias of the update function.
    \item GraphSAGE \cite{graphsage}: proposed a neighbor sampling algorithm for mini-batch GNN training. The model can be specified as follows:
    \begin{equation}
    \begin{split}
        \bm{a}_{v}^{l} & = \bm{h}_{v}^{l-1} || \text{Mean} \left(  \bm{h}_{u}^{l-1}\right) \\
        \bm{h}_{v}^{l} & = \text{ReLU}  \left(\bm{a}_{v}^{l}\bm{W}^{l} + \bm{b}^{l} \right)
    \end{split}
\end{equation}
Where $||$ indicates the concatenation operation.
\end{itemize}
By adopting the aggregate-update paradigm in our system design, our work is capable of training various GNN models.

\subsection{Mini-batch GNN Training}\label{sec:algo}
\begin{figure}[t]
    \centering
    \includegraphics[width=8.5cm]{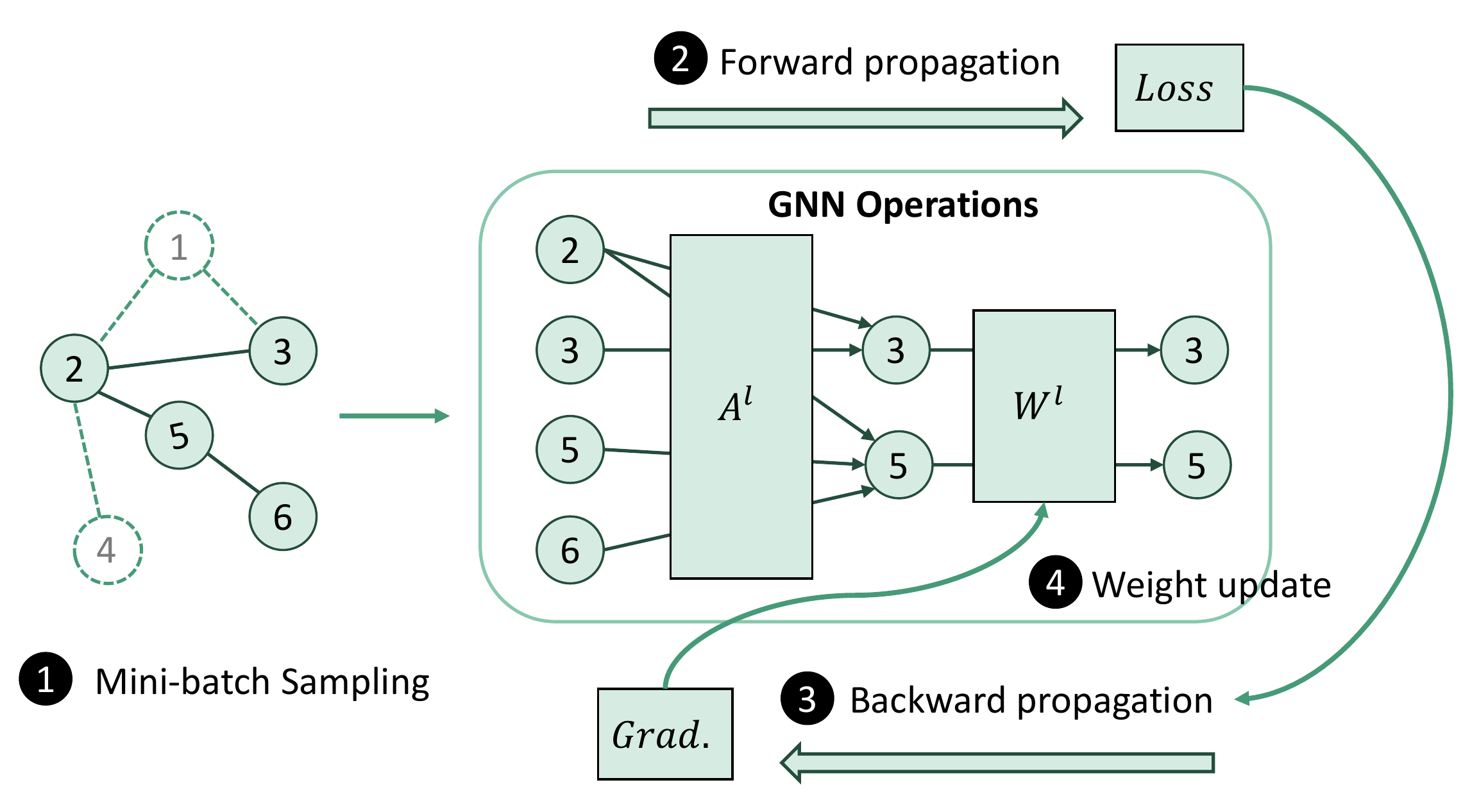}
    \caption{Overview of mini-batch GNN training}
     \label{fig:train_flowl}
\end{figure}

We depict the workflow of mini-batch GNN training in Figure \ref{fig:train_flowl}. 
In each training iteration, a sampler first extracts a mini-batch \{$\mathcal{G}(\mathcal{V}^l,\mathcal{E}^l):1 \leqslant l \leqslant L$\} from the original graph $\mathcal{G}(\mathcal{V},\mathcal{E})$.
The mini-batch serves as a computational graph to perform GNN operations, namely feature aggregation and feature update.
During the forward propagation stage, the GNN operations are performed for $L$ iterations.
The output embeddings $\{\bm{h}_{i}^{L}:v_{i}\in \mathcal{V}^{L}\}$ are compared with the ground truth for loss calculation.
The calculated loss is then used as input for backward propagation.
Backward propagation performs the same set of GNN operations as in forward propagation, but in a reverse direction \cite{understand_GNN};
backward propagation produces the gradients for the weight matrix $\bm{W}^l$ in each layer, which are then used to update the model weights.

\begin{figure}[t]
    \centering
    \includegraphics[width=8.5cm]{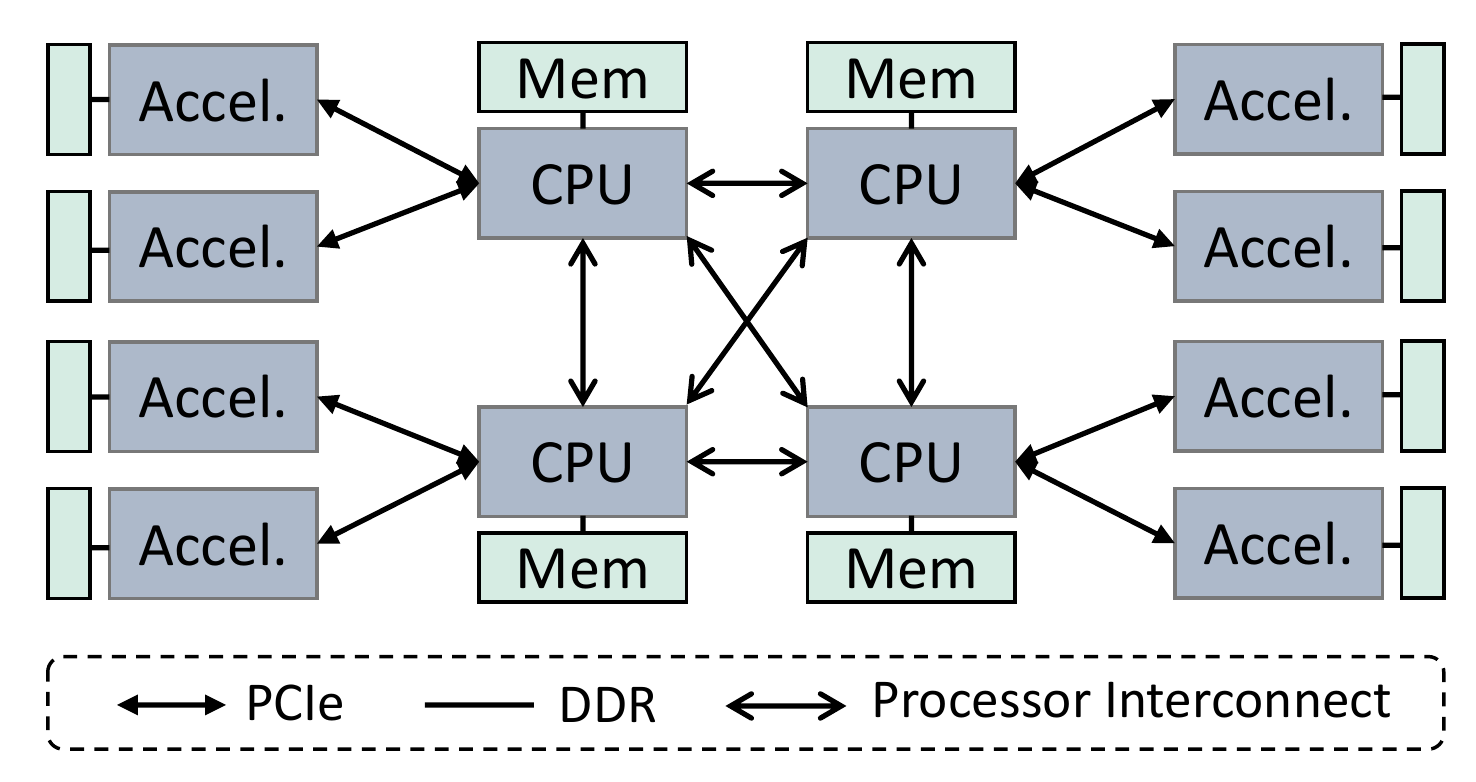}
    \caption{Target heterogeneous architecture}
     \label{fig:platform}
\end{figure}

Our work adopts synchronous Stochastic Gradient Descent (SGD)\cite{sgd} to train GNNs on multiple devices, which performs a similar workflow as the original GNN training but with few variations.
During the first step, multiple mini-batches are sampled and then each device is assigned one mini-batch.
Each device then performs forward/backward propagation as in the original GNN training algorithm.
Finally, the gradients within each device are gathered and averaged. 
The averaged gradients are then broadcast to each device to perform a global weight update.
Training in synchronous SGD on multiple devices is algorithmically equivalent to training with a larger mini-batch on a single device.
For example, training on 4 GPUs with mini-batch size 1024 is equivalent to training on 1 GPU with mini-batch size 4096 \cite{syncsgd}.


\subsection{Target Heterogeneous Architecture}\label{sec:platform}
Figure \ref{fig:platform} shows the target heterogeneous architecture.
It consists of multiple CPUs and multiple accelerators.
The CPU memory on the platform forms a shared address space:
each CPU is able to access the CPU memory to which it is connected, and can also access CPU memory that is connected to other CPUs via processor interconnection channels such as QPI \cite{qpi}.
Each accelerator is connected to a processor via PCIe, and each accelerator is connected to a device memory via DDR channels.




\section{System Design}

In this Section, we first introduce the logical components of HyScale-GNN in Section \ref{sec:hybrid}.
Then, we show how the logical components run on a heterogeneous platform in Section \ref{sec:coord}.
Finally, we introduce the Processor-Accelerator Training Protocol in Section \ref{sec:protocol}, which defines how the processors and the accelerators should interact to perform hybrid training.

\begin{figure}[t]
    \centering
    \includegraphics[width=8.5cm]{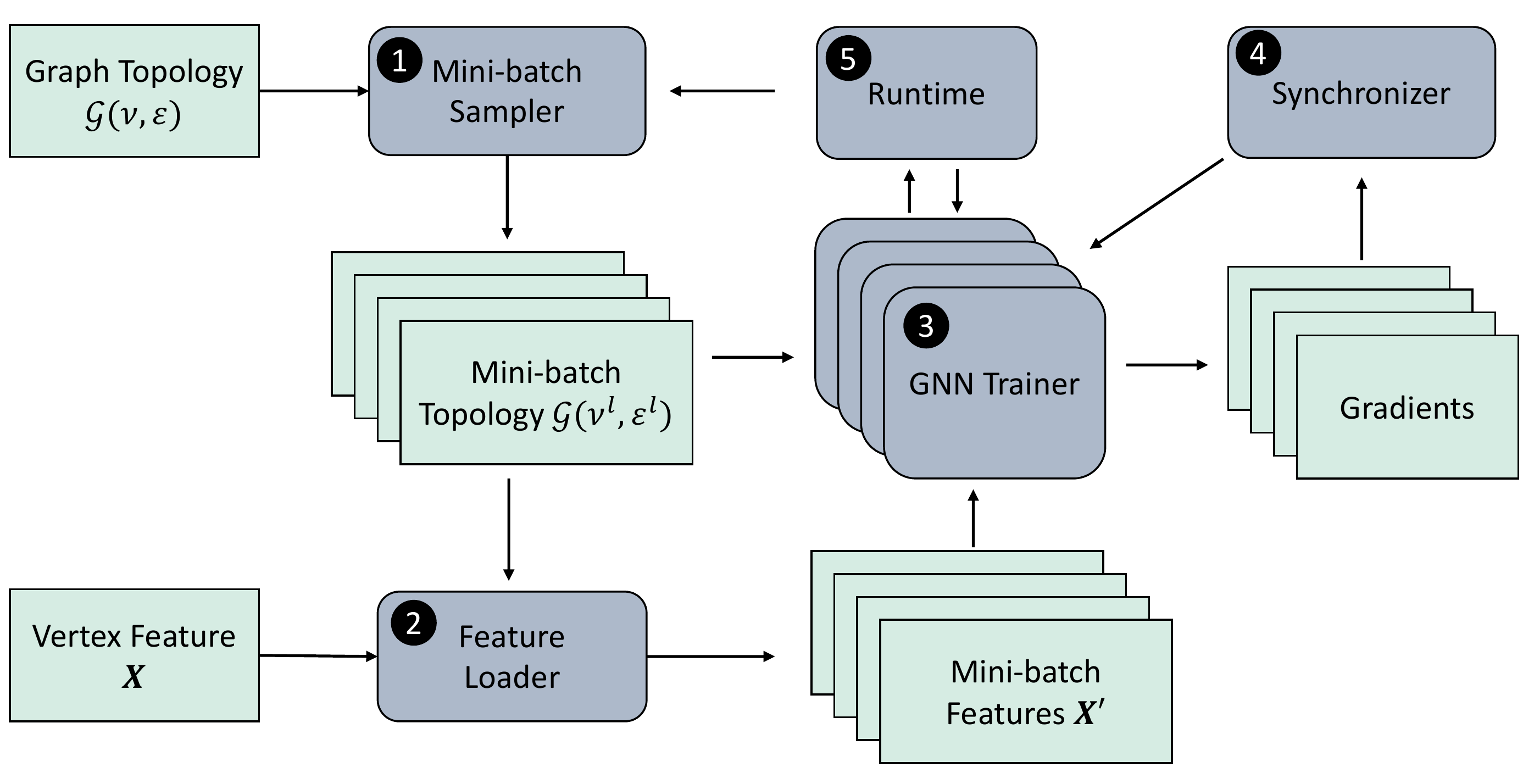}
    \caption{System overview}
     \label{fig:system}
\end{figure} 

\subsection{Hybrid GNN Training System}\label{sec:hybrid}
HyScale-GNN consists of multiple logical components.
We depict the logical components (grey rounded rectangles) and their input/output (green rectangles)  in Figure \ref{fig:system}, and describe each component in the following:

\vspace{0.05cm}
\noindent\textbf{Mini-batch Sampler:} At the beginning of a training iteration, the Mini-batch Sampler takes the graph topology $\mathcal{G}(\mathcal{V},\mathcal{E})$ as input, and produces multiple mini-batches by executing a sampling algorithm \cite{graphsage,graphsaint}.

\vspace{0.05cm}
\noindent\textbf{Feature Loader:}
Given a mini-batch, the Feature Loader extracts a mini-batch feature matrix $\bm{X}'$ from the original feature matrix $\bm{X}$. 
The extracted feature matrix $\bm{X}'$ contains only the vertex features of the sampled vertices instead of all the vertices in the input graph. 


\vspace{0.05cm}
\noindent\textbf{GNN Trainers:}
The GNN Trainers perform the forward propagation and backward propagation of GNN training. 
They take the mini-batch topology and mini-batch feature matrix as inputs, and produce gradients for model weight update.


\vspace{0.05cm}
\noindent\textbf{Synchronizer:}
After each GNN Trainer produces a set of gradients, the Synchronizer performs an all-reduce operation which gathers the gradients from each Trainer, takes the average value of the gradients, and broadcasts the averaged gradients back to each Trainer to update their model weights.

\vspace{0.05cm}
\noindent\textbf{Runtime:}
The Runtime system manages the interaction and data communication between the CPUs and the accelerators based on our Processor-Accelerator Training Protocol (Section \ref{sec:protocol}).
In addition, it also performs Dynamic Resource Management which fine-tunes the workload assignment on the target platform during training.

\begin{figure*}[h]
    \centering
    \includegraphics[width=17cm]{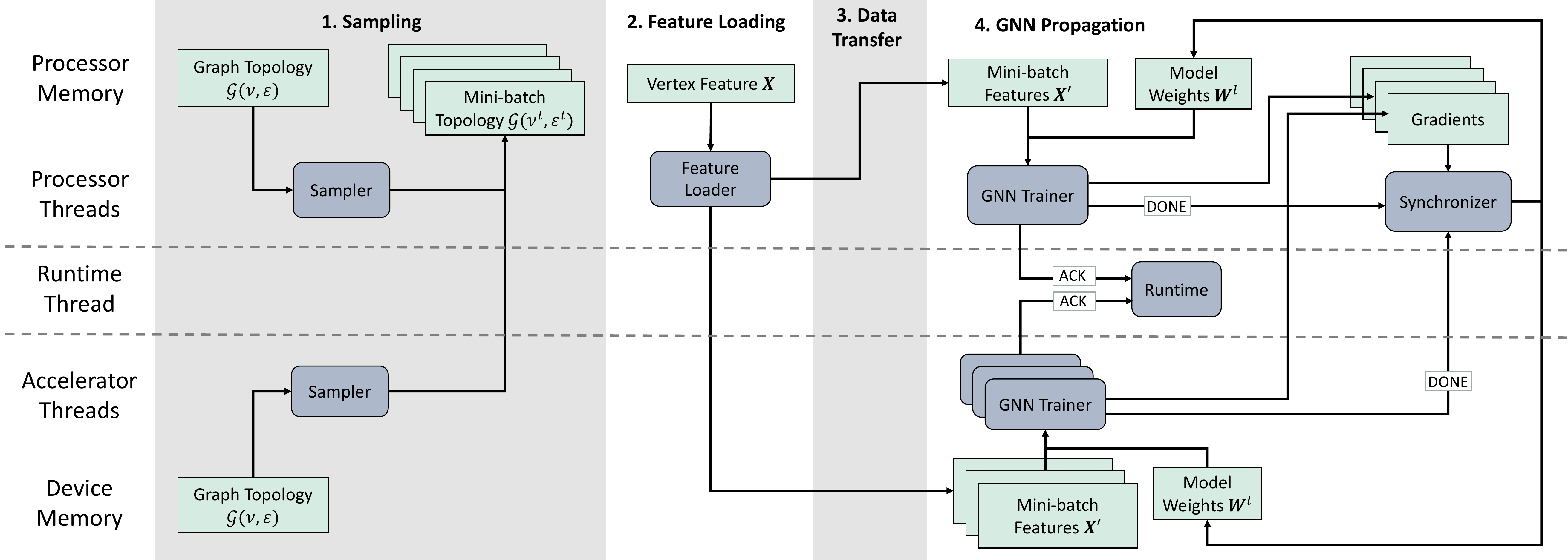}
    \caption{Task mapping and coordination}
     \label{fig:coord}
\end{figure*} 

\begin{figure*}[h]
    \centering
    \includegraphics[width=17cm]{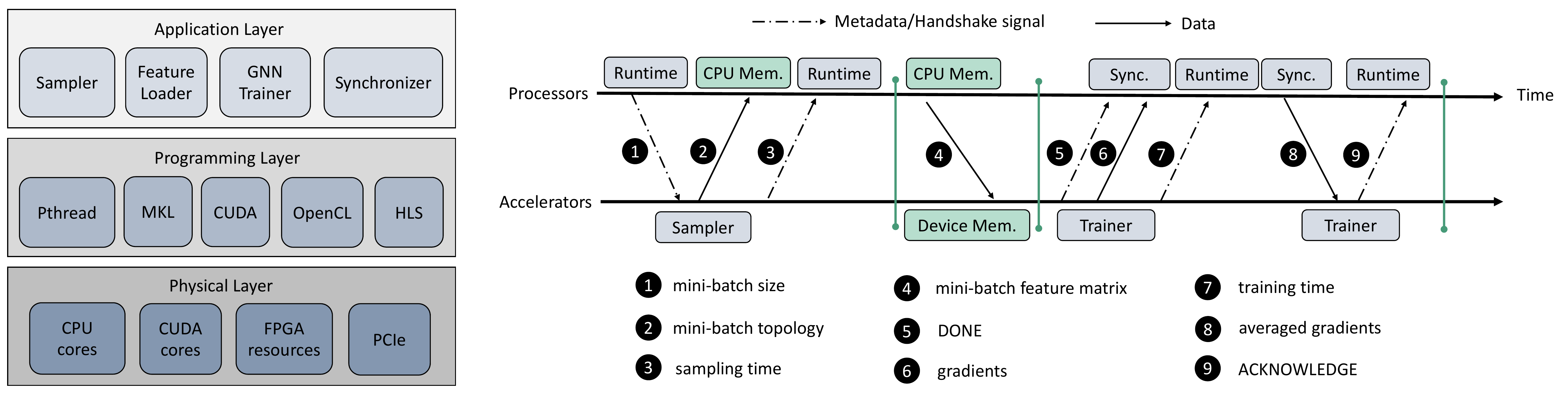}
    \caption{Processor-accelerator training protocol}
     \label{fig:protocol}
\end{figure*} 

\subsection{Task mapping and Coordination}\label{sec:coord}

Our hybrid training system consists of a runtime thread, several processor threads, and several accelerator threads.
Each logical component (Section \ref{sec:hybrid}) is mapped to one or multiple threads.
We show the task mapping and coordination of HyScale-GNN in Figure \ref{fig:coord}.
HyScale-GNN decomposes GNN Training into four pipeline stages: Sampling, Feature Loading, Data Transfer, and GNN Propagation:
(1) \textbf{Sampling}: mini-batch sampling can be performed on both the CPUs and accelerators.
In each training iteration, $n$ mini-batches are produced, where $n$ is the number of GNN Trainers in the system.
After each mini-batch is produced, it is stored in the CPU memory for Feature Loading.
(2) \textbf{Feature Loading}: after collecting all $n$ mini-batches, the Feature Loader reads the feature vector of the sampled vertices from the input feature $\bm{X}$, and stores the loaded features in a sub-matrix $\bm{X}'$.
Feature Loading is only performed on the CPUs. 
This is because an input feature matrix $\bm{X}$ is too large to fit in the device memory for large-scale graphs;
thus, the feature matrix $\bm{X}$ is stored in the CPU memory, and accessed by the Feature Loader which runs on the CPUs.
(3) \textbf{Data Transfer}:
a mini-batch can be either executed on the CPU, or on the accelerator.
If the mini-batch is executed on the accelerator, the mini-batch topology $\{\mathcal{G}(\mathcal{V}^l,\mathcal{E}^l): 1 \leqslant l \leqslant L\}$  and mini-batch feature matrix $\bm{X}'$ are transferred to the accelerator device memory via PCIe.
(4) \textbf{GNN Propagation}:
{in each training iteration, each device (a processor or an accelerator) is assigned a mini-batch topology and a mini-batch feature matrix; these serve as the inputs for the GNN Trainers to perform forward and backward propagation.
Initially, the workload (i.e., mini-batch size) assignment is decided based on our performance model (Section \ref{sec:model}) at design time.
If there is a workload imbalance among the devices at runtime, the DRM engine (Section \ref{sec:drm}) adjusts the workload assignment of the next training iteration.}
After the propagations, each Trainer produces a set of gradients that are later used to update the model weights;
each Trainer then sends a ``DONE" signal to the Synchronizer when the gradients are stored/transferred to the CPU memory.
Since all the accelerators are connected to the CPUs, and the CPUs are connected to each other (Figure \ref{fig:platform}), it is natural to run the Synchronizer on a CPU.
After receiving the ``DONE" signals from all the Trainers, the Synchronizer performs an all-reduce operation, which averages the gathered gradients, and broadcasts the result back to each Trainer.
The Runtime system proceeds to the next training iteration after all the Trainers update their local model weights and send an acknowledgment to the Runtime system.

\subsection{Processor-Accelerator Training Protocol}\label{sec:protocol}
To perform hybrid training on a given heterogeneous architecture, we propose a general \textit{Processor-Accelerator Training Protocol}.
The protocol consists of three layers: the application layer consists of the logical components defined in Section \ref{sec:hybrid};
the programming layer consists of libraries that are used to implement the logical components on multi-core CPUs, GPUs, FPGAs, or AI-specific accelerators;
the physical layer consists of the actual hardware.
HyScale-GNN can be ported to various heterogeneous architectures since the process-accelerator interaction is defined at the application layer, which is not bound to a specific type of accelerator.
We show the data exchange and handshake signals in Figure \ref{fig:protocol}.
Note that Figure \ref{fig:protocol} does not depict the Feature Loading stage since there is no data exchange or handshake signal in that stage.
In each pipeline stage, a barrier is set at the end for synchronization.
In addition, the Runtime system collects the execution time of each stage to fine-tune the workload assignment in the next iteration (Section \ref{sec:drm}).


\section{Optimizations}\label{sec:opt}
In order to achieve high GNN training throughput, we develop various optimizations to perform efficient task mapping (Section \ref{sec:drm}) and to reduce communication overhead (Section \ref{sec:pipe}, \ref{sec:rma}).
It is worth noticing that these optimizations do not alter the semantics of the original GNN training algorithm.
Thus, HyScale-GNN does not trade off the model accuracy and convergence rate for higher training throughput.

\begin{figure*}[h]
    \centering
    \includegraphics[width=17cm]{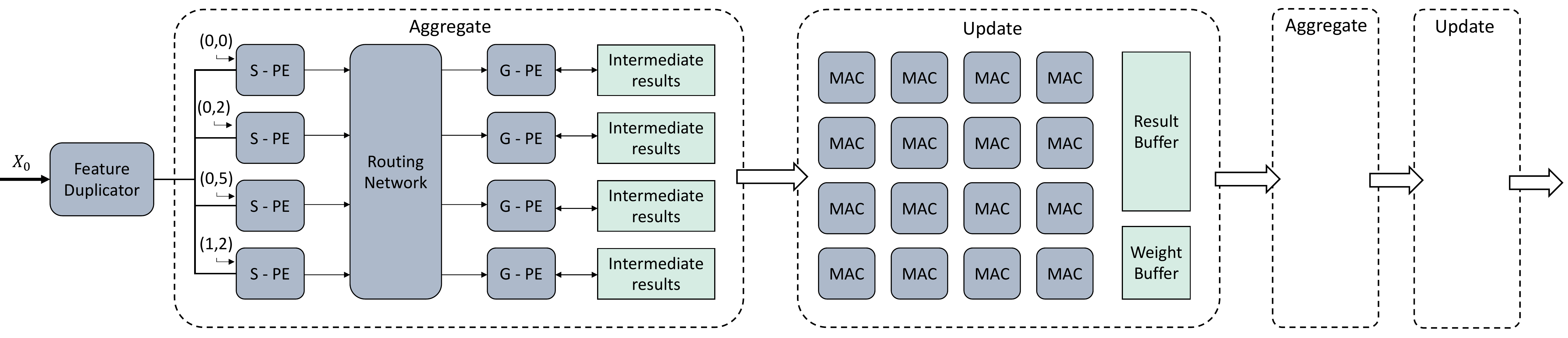}
    \caption{Hardware kernel designs and datapath}
     \label{fig:datapath}
\end{figure*}

\subsection{Dynamic Resource Management}\label{sec:drm}
To efficiently map GNN training tasks onto a heterogeneous architecture,
we first utilize the predicted result from our performance model (Section \ref{sec:model}) to initialize the GNN training task mapping during compile time.
Furthermore, we propose a \textit{Dynamic Resource Management (DRM)} engine that fine-tunes the resource allocation, and task mapping to improve GNN training throughput during runtime.
The DRM engine is a bottleneck-guided optimizer, which improves training throughput by accelerating the bottleneck stage in each iteration.
The DRM engine features two functions to speedup the bottlenecked stage: \textit{balance work} and \textit{balance thread}.
The \textit{balance work} function balances the workload between the CPU and the accelerator by varying the mini-batch size assigned to the Trainers.
The total mini-batch size executed on the hybrid system remains the same after the re-assignment.
The \textit{balance thread} function explores the performance trade-off between CPU tasks (e.g., CPU Sampler, CPU Trainer), and is only used when the bottleneck stage is a CPU task.
It speedups the bottleneck stage by reducing the number of threads assigned to the fastest CPU task, and re-assign those threads to the bottleneck stage.

Algorithm \ref{alg:DRM} describes how the DRM engine works in a high-level view.
First, the DRM engine bundles the Data Transfer time and Training on Accelerator time because the execution time of the two is highly correlated.
For example, if the workload assigned to the accelerator is reduced, the Data Transfer time also reduces since fewer data needs to be communicated.
The DRM engine takes the execution time of each stage as input and identifies the bottleneck stage and the fastest stage.
If the system is bottlenecked by an accelerator task, then the DRM performs \textit{balance work} to adjust the workload assignment between the CPUs and the accelerators in the next iteration.
\begin{small}
\begin{algorithm}[t]
\caption{Dynamic Resource Management}
\label{alg:DRM}
 \textbf{Input}: execution time of Sampling on Accelerator $T_\text{SA}$, Sampling on CPU $T_\text{SC}$, Feature Loading $T_\text{Load}$, Data Transfer  $T_\text{Tran}$, Training on CPU  $T_\text{TC}$, Training on Accelerator $T_\text{TA}$ \\
 \textbf{Output}: thread assignment, workload assignment
\begin{algorithmic}[1]
\State{$T_\text{Accel} = \textbf{max(} T_\text{Tran}, T_\text{TA} \textbf{)}$}
\State{Sorted\_all = \textbf{sort(}$T_\text{SC},T_\text{SA},T_\text{Load}, T_\text{TC}, T_\text{Accel} $\textbf{)}} 
\State{fastest = Sorted\_list[4]}
\State{second = Sorted\_list[3]}
\State{bottleneck = Sorted\_list[0]}
\State{}
\State{Sorted\_CPU = \textbf{sort(}$T_\text{SC},T_\text{Load}, T_\text{TC}$)}
\State{fastest\_CPU\_task = Sorted\_CPU[2]}
\State{}
\Switch{bottleneck}
{\color{blue}\Comment{Bottleneck-guided Optimizer}}
\Case{$T_\text{SA}:$}
\State{\textbf{balance\_work}($T_\text{SC}$, $T_\text{SA}$)}
\EndCase
\Case{$T_\text{Accel}:$}
\State{\textbf{balance\_work}($T_\text{TC}$, $T_\text{Accel}$)}
\EndCase
\Case{$T_\text{Load}:$}
\State{\textbf{balance\_thread}(fastest\_CPU\_task, bottleneck)}
\EndCase
\Case{$T_\text{SC}:$}
\If{fastest == $T_\text{SA}$}
\State{\textbf{balance\_work}($T_\text{SC}$, $T_\text{SA}$)}
\ElsIf{(fastest == $T_\text{Accel}$ and second == $T_\text{SA}$)}
\State{\textbf{balance\_work}($T_\text{SC}$, $T_\text{SA}$)}
\Else
\State{\textbf{balance\_thread}(fastest, bottleneck)}
\EndIf
\EndCase
\Case{$T_\text{TC}:$}
\If{fastest == $T_\text{Accel}$}
\State{\textbf{balance\_work}($T_\text{TC}$, $T_\text{Accel}$)}
\ElsIf{(fastest == $T_\text{SA}$ and second == $T_\text{Accel}$)}
\State{\textbf{balance\_work}($T_\text{TC}$, $T_\text{Accel}$)}
\Else
\State{\textbf{balance\_thread}(fastest, bottleneck)}
\EndIf
\EndCase
\EndSwitch
\end{algorithmic}
\end{algorithm}
\end{small}
If the system is bottlenecked by the Feature Loader, the DRM engine performs \textit{balance thread} which re-allocates more threads to perform Feature Loading.
If the system is bottlenecked by the CPU Sampler, the DRM engine can either perform \textit{balance work} or \textit{balance thread} to speedup the CPU Sampler.
The decision depends on which stage runs the fastest.
If the Accelerator Sampler runs the fastest, the DRM engine performs \textit{balance work} which increases the workload assigned to the accelerators;
if the fastest task is the Accelerator Trainer, followed by the Accelerator Sampler, then the DRM engine also assigns more workload to the accelerators; 
otherwise, the DRM engine performs \textit{balance thread} which explores performance trade-offs between other CPU tasks.
Finally, if the system is bottlenecked by the CPU Trainer, the DRM engine can also improve the performance by performing either \textit{balance work} or \textit{balance thread};
thus, the improvement strategy is similar to when bottlenecked by the CPU Sampler.

\begin{figure}[t]
    \centering
    \includegraphics[width=8.7cm]{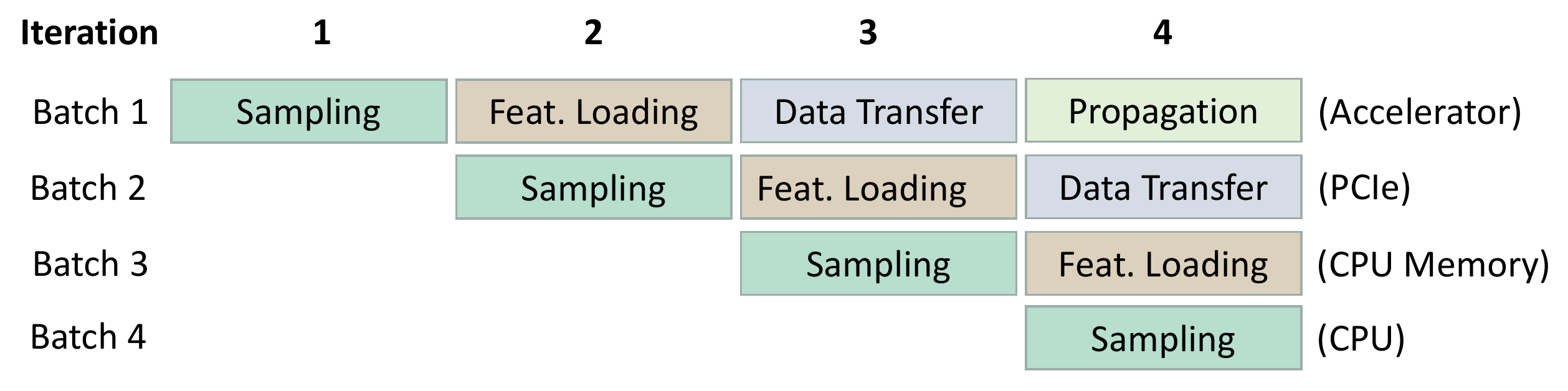}
    \caption{Two-stage Feature Prefetching}
     \label{fig:pipeline}
\end{figure} 

\subsection{Two-stage Feature Prefetching}\label{sec:pipe}
HyScale-GNN runs GNN training on both the CPUs and the accelerators.
For GNN Trainers that run on the accelerators, the data needs to be fetched from the CPU memory and then transferred to the accelerators via PCIe. 
To reduce the expensive communication overhead, HyScale-GNN performs Feature Prefetching.
The idea is to pre-load the mini-batches for the next training iteration onto the accelerators first, so the accelerators do not have to wait for data when performing GNN operations.
Observing that the Feature Prefetching stage can still bottleneck the system, we further decompose Feature Prefetching into two pipeline stages: Feature Loading, and Data Transfer.
The Feature Loading stage loads the Mini-batch Feature $\bm{X}'$ from the CPU memory, and the Data Transfer stage sends the matrix $\bm{X}'$ to the accelerator via PCIe.
The two stages can run concurrently because they utilize different memory channels (CPU RAM channel, and PCIe channel), and there is no data dependency between mini-batches.
Figure \ref{fig:pipeline} shows an example of the Two-stage Feature Prefetching: in iteration 4, while an accelerator is executing mini-batch 1, the feature matrix of mini-batch 2 is being transferred to the accelerator via PCIe, and the feature matrix of mini-batch 3 is being loaded from the CPU memory, simultaneously. 
Thus, when the accelerator executes mini-batch 2 in the next iteration, the mini-batch topology and mini-batch features are ready in the accelerator's device memory.
Note that Figure \ref{fig:pipeline} only shows a simplified version of the system pipeline.
For each iteration, multiple mini-batches can be sampled, loaded, transferred, and executed.
It is also worth noting that our system pipeline efficiently utilizes the various resources on the heterogeneous architecture.

\subsection{Hardware Kernel Design}\label{sec:rma}
GNN training incurs a massive amount of memory access, and the expensive memory access overhead limits the training throughput.
Thus, we design dedicated hardware kernels and datapath to reduce external memory access for the GNN Propagation stage as shown in Figure \ref{fig:datapath}. 
GNN propagation consists of an aggregation stage and an update stage (Section \ref{sec:GNN}). 
For the update stage, we adopt a systolic-array-based design to perform Multi-Layer Perceptron (MLP); for the aggregation stage, we adopt a scatter-gather design \cite{hygcn,hp-gnn} to process multiple edges in parallel.
Figure \ref{fig:datapath} shows an example of a kernel with four sets of scatter-gather processing elements (PEs) which can process four edges in parallel.
To maximize data reuse, HyScale-GNN first sorts the edges within a mini-batch by their source vertex so that edges with the same source vertex are executed in a back-to-back manner.
When a vertex feature is fetched from the external memory, the Feature Duplicator broadcasts the fetched feature to all the scatter-PEs (S-PEs).
The feature is then stored in the local memory of the S-PEs and reused $D_{out}(v)$ times where $D_{out}(v)$ is the out-degree of vertex $v$.
For example, in Figure \ref{fig:datapath}, four edges are processed.
Assume $D_{out}(v_0)$ is 3, then the loaded feature $X_0$ can be reused three times at most.
Since the edges are sorted by source vertex, the first three edges have the same source vertex $v_0$, and $X_0$ is reused three times.
The fourth S-PE remains idle until $X_1$ is read from memory.
This design maximizes the input data reuse since each vertex feature only needs to be read once from memory, and the input memory traffic is reduced from $O(\mathcal{E}^1)$ to $O(\mathcal{V}^0)$ (notations are defined in Table \ref{tab:notations}).
To reduce memory footprint, we design a customized datapath, which avoids writing the intermediate results back to the memory during GNN Propagation.
As shown in Figure \ref{fig:datapath}, the output of the aggregate kernel is directly sent to the update kernel,
and the output of the update kernel is sent to the aggregate kernel for feature aggregation in the next layer.
Therefore, only the final output is written back to the memory.


\section{Performance Model}\label{sec:model}
To initialize the task mapping on a heterogeneous architecture, we propose a performance model to predict the performance of HyScale-GNN.
First, we define the GNN training throughput as million of traversed edges per second (MTEPS):
 \begin{equation}
   \text{Throughput} = \frac{\sum_{i=0}^{n}\sum_{l=1}^L|\mathcal{E}_i^l|}{T_{\text{execution}}}
    \label{eq:throughput}
\end{equation}
We use $n$ to denote the number of GNN Trainers running on the system.
Each Trainer executes one mini-batch in each iteration.
Therefore, the numerator denotes the total number of edges traversed by all the Trainers in one iteration, and the denominator denotes the execution time of one training iteration (Section \ref{sec:algo}).
There are four pipeline stages in our system (Section \ref{sec:hybrid}): Sampling, Feature Loading, Data Transfer, and GNN Propagation;
thus, $T_{\text{execution}}$ is modeled as:
 \begin{equation}
   T_{\text{execution}} = \max(T_{\text{samp}},T_{\text{load}},T_{\text{trans}},T_{\text{prop}})
    \label{eq:tex}
\end{equation}
Instead of formulating an equation, we estimate $T_{\text{samp}}$ by running the sampling algorithm under different numbers of threads and different mini-batch sizes, and deriving their execution time during design phase. 
This is because the computation pattern varies in different sampling algorithms \cite{graphsage,graphsaint}, so it is not feasible to model the sampling time $T_{\text{samp}}$ of various algorithms with a simple equation.

$T_{\text{load}} \text{ and } T_{\text{tran}}$ can be modeled as:
 \begin{equation}
   T_{\text{load}} = \frac{\sum_{i=0}^{n}|\mathcal{V}^0|\times f^0 \times S_\text{feat}}{BW_\text{DDR}}
    \label{eq:tload}
\end{equation}
 \begin{equation}
   T_{\text{trans}} = \frac{|\mathcal{V}^0|\times f^0 \times S_\text{feat}}{BW_\text{PCIe}}
    \label{eq:ttran}
\end{equation}
The numerator in Equation \ref{eq:tload} denotes the size of vertex features that need to be loaded from the CPU memory, and the numerator in Equation \ref{eq:ttran} denotes the size of vertex features that need to be transferred to a single accelerator.
$S_\text{feat}$ denotes the data size, which is 4 bytes for a single-precision floating-point.
For $T_{\text{load}}$, the data is loaded from the CPU memory, so the denominator is the CPU memory bandwidth; for $T_{\text{tran}}$, the denominator is the PCIe bandwidth.
We use ``bandwidth" to denote the effective bandwidth of performing burst data transactions as opposed to the peak bandwidth.

Multiple GNN Trainers run in parallel, and $T_{\text{train}}$ can be modeled as:
\begin{equation}
T_{\text{prop}} = \max_{i \in n}(T_{\text{Trainer,}i}) + T_{\text{sync}}
    \label{eq:ttrain}
\end{equation}
The execution time of a single Trainer can be modeled as:
\begin{equation}
\begin{split}
    T_{\text{Trainer}} = t_\text{forward\_prop} + t_\text{backward\_prop} = \\ 
    \sum_{l=1}^L \oplus(t_{\text{aggregate}}^l,t_{\text{update}}^l) + \\ t_{\text{update}}^{1} + \sum_{l=2}^L \oplus(t_{\text{aggregate}}^l,t_{\text{update}}^l) 
\end{split}
    \label{eq:ttrainer}
\end{equation}
which is the total time to perform forward propagation and backward propagation.
The $\oplus$ operator depends on the kernel design. 
If feature aggregation and feature update are pipelined (e.g., Trainer on FPGA), then $\oplus$ is the $\max$ operator;
if they are not pipelined (e.g., Trainer on CPU), then $\oplus$ is the $\sum$ operator.
$t_{\text{aggregate}}^l \text{ and }t_{\text{update}}^l$ can be modeled as:
 \begin{equation}
    t_{\text{aggregate}}^l  = \frac{|\mathcal{E}^{l-1}|\times f^{l} \times S_{\text{feat}}}{BW_\text{DDR} }
    \label{eq:agg}
\end{equation}
\begin{equation}
    t_{\text{update}}^l = \frac{|\mathcal{V}^l|\times f^l \times f^{l+1}}{N \times {freq.}} 
    \label{eq:ns_update}
\end{equation}
The aggregation time $t_{\text{aggregate}}^l$ is modeled as the traffic size of fetching the feature vector of the source vertices, divided by the memory bandwidth.
The memory bandwidth depends on where the Trainer is located: for the CPU Trainer, it fetches data from the CPU memory; for Accelerator Trainer, it fetches data from the device memory.
Since $|\mathcal{E}^l|$ edges are processed during the $l$-th layer feature aggregation, the traffic size can be modeled as ${|\mathcal{E}^{l-1}|\times f^{l} \times S_{\text{feat}}}$.
The update time $t_{\text{update}}^l$ is modeled as the number of multiply-and-add (MAC) operations that are performed during feature update, divided by the computing power of the GNN Trainer.
We model the computing power as  $N \times {freq.}$ where $N$ is the number of computation parallelism (e.g., MAC units) in each trainer, and $freq.$ is the operating frequency.
$T_{sync}$ can be model as:
 \begin{equation}
   T_{\text{sync}} = \frac{\sum_{l=1}^L f^{l-1} \times f^l \times S_\text{feat}}{BW_\text{PCIe}} \times 2
    \label{eq:tsync}
\end{equation}
The numerator is the model size (i.e., total size of all of the weight matrices).
The factor of 2 comes from the all-reduce operation where the model is first gathered, averaged, and then scattered back to each Trainer;
thus, the data is transferred through the PCIe twice.
The denominator is the PCIe bandwidth.

\section{Experimental Results}

\subsection{Experimental Setup}
\subsubsection{Environment}
We conduct our experiments on a dual-socket server, which consists of two AMD EPYC 7763 CPUs.
We evaluate HyScale-GNN using two heterogeneous setups: a CPU-GPU heterogeneous architecture, and a CPU-FPGA heterogeneous architecture.
For the CPU-GPU heterogeneous architecture, the dual-socket server is connected to four Nvidia A5000 GPUs;
for the CPU-FPGA heterogeneous architecture, the dual-socket server is connected to four Xilinx U250 FPGAs.
We list the detailed specification of the CPU, GPU, and FPGA in Table \ref{tab:spec}.
We implement the multi-GPU baseline, and CPU-GPU design using Python v3.8, PyTorch v1.11, CUDA v11.3, and PyTorch-Geometric v2.0.3.
We develop our FPGA kernels using Xilinx Vitis HLS v2021.2 \cite{vitis}.

\begin{table}[h]
\centering
\caption{Specifications of the platforms }
\begin{threeparttable}
 \begin{adjustbox}{max width=0.98\columnwidth}
\renewcommand{\arraystretch}{1.05}
\begin{tabular}{c|c|c|c}
 \toprule
\textbf{Platforms} & \begin{tabular}[|c|]{@{}c@{}} CPU \\  AMD EPYC 7763 \end{tabular}  & \begin{tabular}[|c|]{@{}c@{}} GPU \\  Nvidia RTX A5000 \end{tabular} & \begin{tabular}[|c|]{@{}c@{}} FPGA \\  Xilinx Alveo U250 \end{tabular}  \\ 
\midrule \midrule
 {Technology}  & TSMC 7 nm+   & Samsung 8 nm & TSMC 16 nm \\ 
{Frequency} & 2.45 GHz  & 2000 MHz & 300 MHz 
      \\ 
{Peak Performance}& 3.6 TFLOPS & 27.8 TFLOPS & 0.6 TFLOPS  \\ 
{On-chip Memory}& 256 MB L3 cache & 6 MB L2 Cache & 54 MB  \\
{Memory Bandwidth}& 205 GB/s & 768 GB/s & 77 GB/s   \\ \bottomrule
\end{tabular}
\end{adjustbox}
\end{threeparttable}
\label{tab:spec}
\end{table}

\begin{small}
\begin{table}[t]
\renewcommand{\arraystretch}{1.1}
\caption{Statistics of the Datasets and GNN-layer dimensions}
    \centering
    \begin{tabularx}{0.98\columnwidth}{cccXXX}
        \toprule
        \textbf{Dataset} & \textbf{\#Vertices} & \textbf{\#Edges} & $f_{0}$ &  $f_{1}$ &  $f_{2}$\\
        \midrule
        \midrule
        ogbn-products  & 2,449,029 & 61,859,140 &  100 & 256 & 47  \\
        ogbn-papers100M & 111,059,956 & 1,615,685,872 &  128 & 256 & 172\\
        MAG240M (homo) & 121,751,666 & 1,297,748,926 &  756 & 256 & 153\\
        \bottomrule
    \end{tabularx}
    \label{tab: graph-scale}
\end{table}
\end{small}

\begin{table}[h]
\renewcommand{\arraystretch}{1}
\caption{Hardware Parameters and Resource utilization}
    \centering
    
    \begin{tabularx}{0.9\columnwidth}{cXXXX}
        \toprule
        {Parallelism (\textit{n,m})} & LUTs & DSPs & URAM & BRAM   \\
        \midrule \midrule
        (8, 2048) & 72\% & 90\% & 48\% & 40\% \\ 
        \bottomrule
    \end{tabularx}

    \label{tab: resource}
\end{table}

\subsubsection{GNN Models and Datasets}
We choose two widely used GNN models: GCN \cite{gcn}, and GraphSAGE \cite{graphsage} to evaluate our system.
We adopt a commonly used model setup: a two-layer model with a hidden feature size of 256.
We choose a medium-scale dataset, and two large-scale datasets with over billion edges for evaluation: ogbn-products, ogbn-papers100M \cite{ogb}, and  MAG-240M (homo) \cite{hu2021ogblsc}.
The ogbn-products dataset is a medium-scale graph with 60 million edges; we include this dataset to compare our performance with previous works.
The MAG-240M (homo) is the homogeneous version of the MAG-240M dataset, which only preserves one type of vertex and one type of edge in the original heterogeneous graph.
Note that MAG-240M (homo) still contains 1.3 billion edges, making it a large-scale graph.
Details of the datasets and the GNN-layer dimensions are shown in Table \ref{tab: graph-scale}.
We use the Neighbor Sampler \cite{graphsage} to produce mini-batches;
we set the mini-batch size as 1024, and the neighbor sampling size of each layer is 25 and 10.

\subsection{System Implementation}

We show how the Processor-Accelerator Training Protocol is implemented using the libraries in the programming layer (Section \ref{sec:protocol}) in Listing \ref{lst:impl}; {while we use GPU and FPGA as examples, the processor-accelerator interaction is similar if the protocol is adapted to other AI-accelerators.}
We implement the Runtime system using Pthreads.
We launch multiple threads to exchange data, handshake, or launch accelerator kernels.
Data transfer and kernel launching can be realized using APIs provided by the programming libraries such as CUDA and OpenCL.
To implement the handshake, we utilize the \textit{condition wait} function in Pthreads.
For example, the Synchronizer needs to wait for all the Trainers to complete GNN propagation before averaging the gradients.
Each Trainer increments the ``DONE" variable upon producing the gradients and then prompts the synchronizer.
When ``DONE" equals the number of Trainers, the Synchronizer proceeds to average the gathered gradients.

{We list the hardware parameters and resource utilization of the FPGA design in Table \ref{tab: resource}.
We use $n$ and $m$ to denote the parallelism of the aggregate kernel and update kernel, respectively. 
In particular, $n$ indicates the number of scatter-gather PEs. 
$m$ indicates the number of multiply-and-accumulate units in the systolic-array-based kernel design. Figure \ref{fig:datapath} shows an example for $n=4$ and $m=16$.}
\begin{lstlisting}[float=tp,language=C, caption=System implementation, label={lst:impl}]
//send data to the accelerator
q.enqueueMigrateMemObjects(input, 0) //FPGA

cudaMemcpy(gpu_input, input, data_size, cudaMemcpyHostToDevice); //GPU

//read data from the accelerator
q.enqueueMigrateMemObjects(result, CL_MIGRATE_MEM_OBJECT_HOST)); //FPGA

cudaMemcpy(result, gpu_result, res_size, cudaMemcpyDeviceToHost); //GPU

//launch an accelerator kernel
q.enqueueTask(gnn_krnl, NULL, &event)); //FPGA

gnn_krnl<<<dimGrid, dimBlock>>>(input);	//GPU

//get execution time on the accelerator
start = getProfilingInfo(PROFILING_COMMAND_START);
end = getProfilingInfo(PROFILING_COMMAND_END);
time = end - start;

//handshake
Synchronizer_thread:
    pthread_mutex_lock(&mutex);
    while (DONE != n) //n is the number of Trainers
        pthread_cond_wait(&cond, &mutex);
    gather_data();
    average_gradients();
    pthread_mutex_unlock(&mutex);

Trainer_threads:
    GNN_training(); //CPU function, or launch an accelerator kernel
    pthread_mutex_lock(&mutex);
    DONE++
    pthread_cond_signal(&cond);
    pthread_mutex_unlock(&mutex);
    
\end{lstlisting}

\subsection{Evaluation of Performance Model}\label{sec:model_ev}
\begin{figure}[t]
    \centering
    \includegraphics[width=8cm]{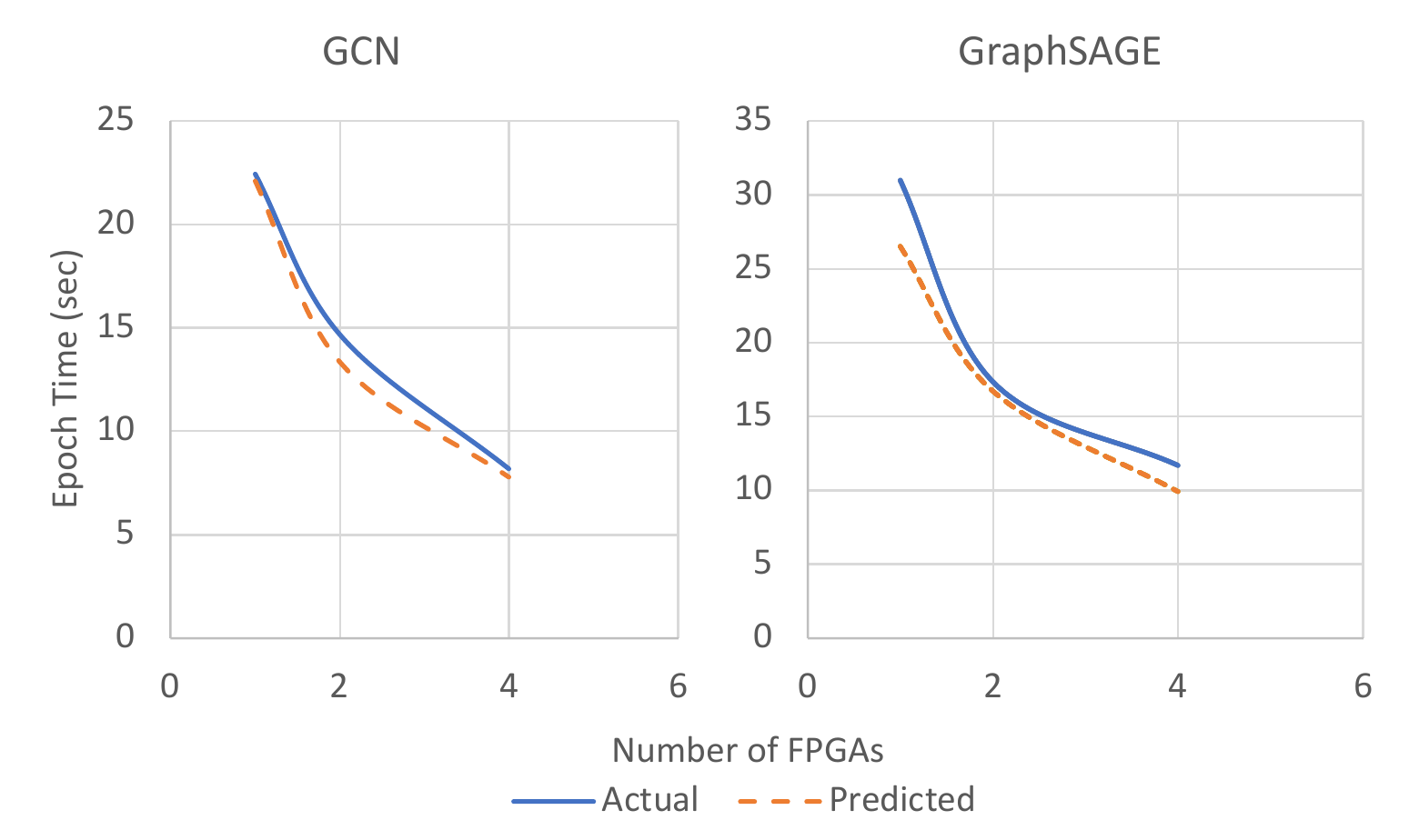}
    \caption{Predicted performance vs. actual performance on two GNN models}
     \label{fig:model}
\end{figure}

We evaluate our performance model by comparing the predicted epoch time with the actual experimental result.
Figure \ref{fig:model} shows the epoch time comparison on the MAG240M (homo) dataset under various number of FPGAs.
The prediction error ranges from 5\% to 14\% on average.
The error comes from extra latency that is not formulated in our model.
First, there is an initial overhead when launching the kernel on an accelerator.
Second, the overhead of pipeline flushing \cite{tgnn} is not included in the model.
These two overheads are hard to predict as they depend on various factors such as the target accelerator and the version of the compiler.

\begin{figure}[t]
    \centering
    \includegraphics[width=7.8cm]{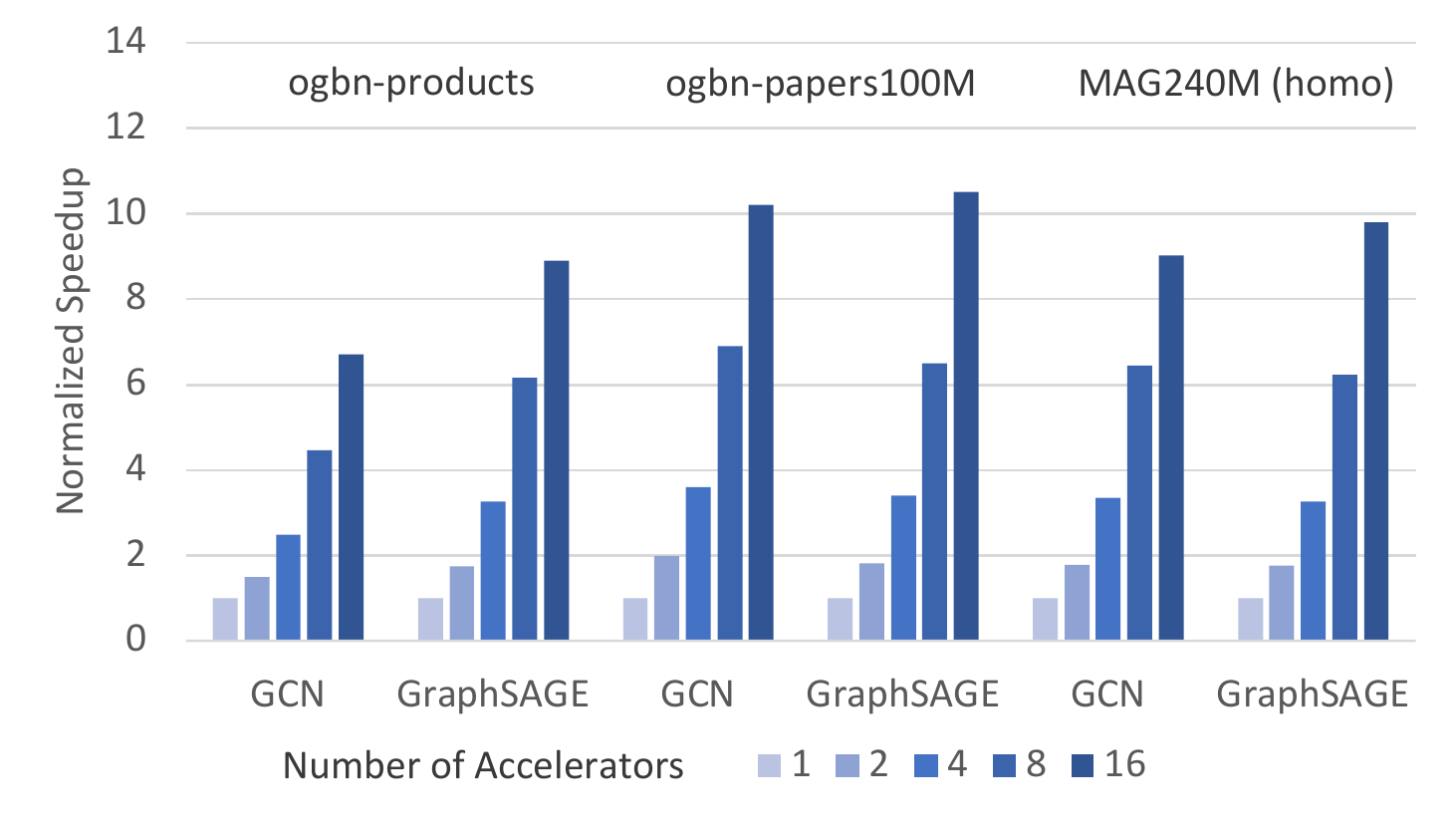}
    \caption{Scalability of our hybrid training system}
     \label{fig:scale}
\end{figure}

\subsection{Scalability}
We evaluate the scalability of HyScale-GNN using our performance model (Section \ref{sec:model}).
We show the scalability of HyScale-GNN in Figure \ref{fig:scale}.
Using the CPU-FPGA platform as an example, HyScale-GNN demonstrates good scalability to 16 FPGAs.
The limiting factor of scalability is the CPU memory bandwidth.
As we use more accelerators, more mini-batch feature matrices need to be loaded from the CPU memory.
We observe that the CPU memory starts to saturate when more than 12 accelerators are used on the heterogeneous platform.
The first set of experiments which runs a GCN model on the ogbn-products dataset shows lower scalability than other sets of experiments.
This is because the first set of experiments is bottlenecked by the data transfer time (i.e., PCIe bandwidth), which limits the amount of workload that can be assigned to the accelerators and thus limits the achievable speedup.

\subsection{Overall Performance}

\subsubsection{Performance evaluation}\label{sec:perf_ev}
\begin{figure}[t]
    \centering
    \includegraphics[width=8.6cm]{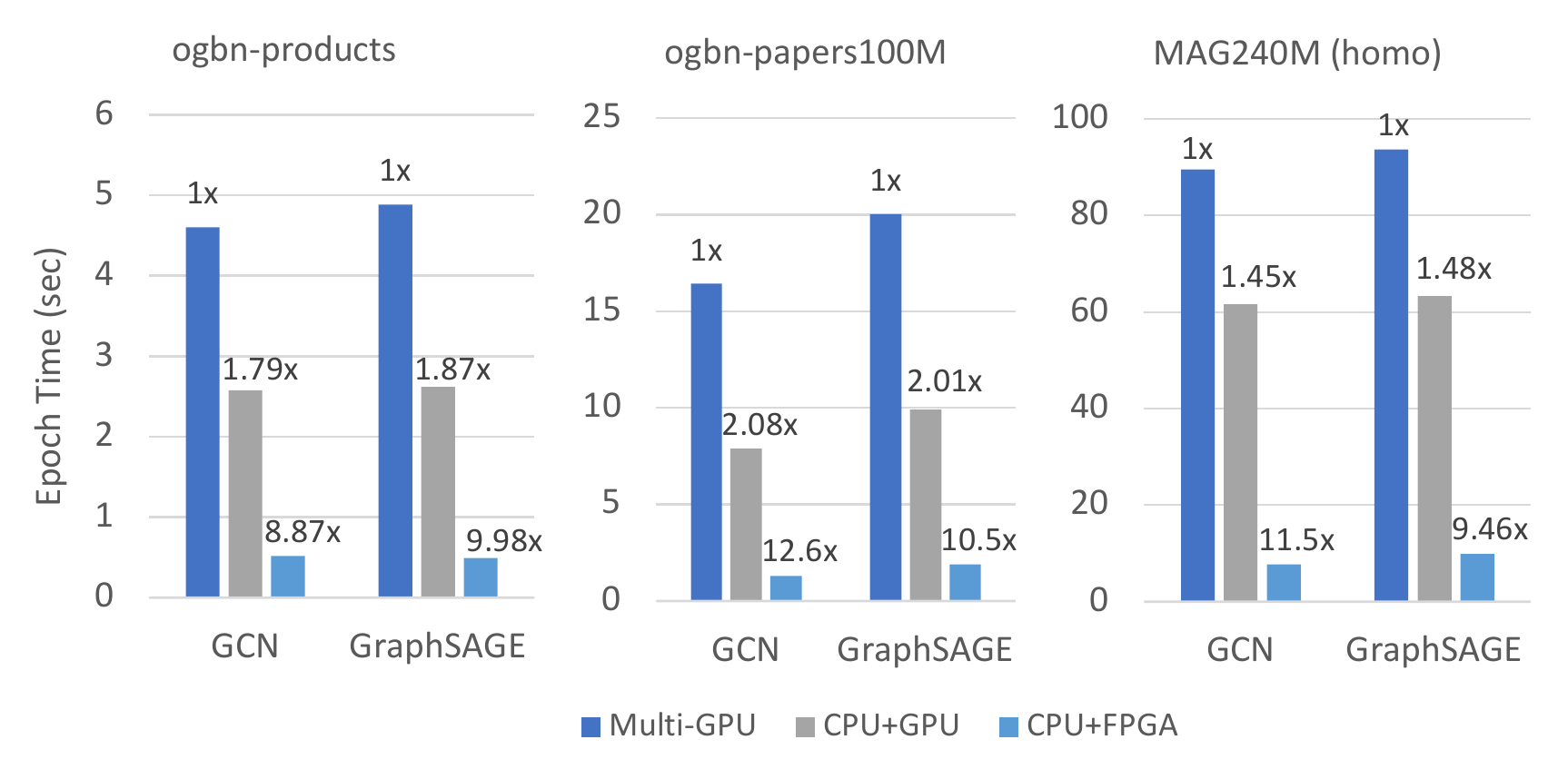}
    \caption{Cross platform comparison}
     \label{fig:perf}
\end{figure}

We evaluate the performance of HyScale-GNN using a CPU-GPU heterogeneous architecture, and a CPU-FPGA heterogeneous architecture.
We compare the epoch time of HyScale-GNN with a state-of-the-art multi-GPU GNN training implementation using PyTorch-Geometric (PyG) \cite{pyg}.
The PyG baseline also runs on the CPU-GPU heterogeneous architecture; 
however, it does not utilize the CPU to perform hybrid training, so we regard it as a multi-GPU baseline.
We show the result in Figure \ref{fig:perf}.
By applying various optimizations and performing hybrid CPU-GPU training, HyScale-GNN achieves up to $2.08\times$ speedup compared with the multi-GPU baseline.
We discuss the effectiveness of each optimization in Section \ref{sec:ablation}.
On the CPU-FPGA heterogeneous architecture, HyScale-GNN achieves up to $12.6\times$ speedup compared with the multi-GPU baseline, and $5\times -6\times$ speedup compared with the CPU-GPU heterogeneous architecture.
This is because FPGAs feature customizable datapath and memory organization, which allows the Accelerator Trainer to minimize external memory access during GNN training.
In particular, all the intermediate results are stored on-chip using the abundant on-chip memory of FPGA, and only the final result is written back to the memory.
In contrast, GPUs suffer from frequent memory access throughout the training since traditional cache policies fail to capture the data access pattern in GNN training \cite{cachemiss}.

\subsubsection{Comparison with State-of-the-art}
Many works \cite{graphact,hp-gnn,pyg,rubik,dglv2,distdgl,pagraph,p3} have been proposed to accelerate GNN training.
However, only a few of the works are capable of training GNN models on large-scale graphs. 
We choose three representative GNN training systems for comparison, namely PaGraph \cite{pagraph}, $P^3$ \cite{p3}, and DistDGLv2 \cite{dglv2}.
We list the platform setup of each work in Table \ref{tab:sota}.
We use \textit{SAGE} to indicate the GraphSAGE \cite{graphsage} model.
Among the three large-scale GNN training systems, PaGraph is the only work that runs on a single node;
$P^3$ and DistDGLv2 run on a distributed platform with four nodes and eight nodes, respectively.
\begin{table}[t]
\renewcommand{\arraystretch}{1.1}
\caption{Platform Setup of State-of-the-art}
\label{tab:sota}
\begin{adjustbox}{max width=\columnwidth}
\begin{tabular}{lcccc}
\toprule
            & \begin{tabular}[c]{@{}c@{}}number of \\compute node(s)\end{tabular} & Setup of each node                                                           & Sample size & \begin{tabular}[c]{@{}c@{}}Hidden\\ dim.\end{tabular} \\ \hline \hline
PaGraph \cite{pagraph}     & 1                                                      & \begin{tabular}[c]{@{}c@{}}2 Xeon Platinum 8163\\ 8 Nvidia V100\end{tabular} & (25, 10)    & 256                                                        \\ \hline
$P^3$ \cite{p3}         & 4                                                      & \begin{tabular}[c]{@{}c@{}}1 Xeon E5-2690\\ 4 Nvidia P100 (2016) \end{tabular}       & (25, 10)    & 32                                                         \\ \hline
DistDGLv2 \cite{dglv2} & 8                                                      & \begin{tabular}[c]{@{}c@{}}96 vCPU\\ 8 Nvidia T4\end{tabular}                & (15, 10, 5) & 256                                                        \\ \hline
This Work   & 1                                                      & \begin{tabular}[c]{@{}c@{}}2 EPYC 7763\\ 4 Xilinx U250\end{tabular}          & -           & -                                                          \\ \bottomrule
\end{tabular}
\end{adjustbox}
\end{table}

We compare the epoch time of our work, which runs on a single node using only 4 FPGAs, with the aforementioned training systems.
For each set of experiments, we set the same model configuration (sample size, hidden dimension) as the work we are comparing with.
As shown in Table \ref{tab:sota_perf}, HyScale-GNN achieves up to $6.9\times$ and $5.27\times$ speedup compared with PaGraph and $P^3$, respectively.
{To provide a fair comparison, we normalize the epoch time w.r.t. platform peak performance; this metric shows the effectiveness and efficiency of the system design itself, rather than relying on powerful hardware to deliver high performance. As shown in Table \ref{tab:norm_perf}, HyScale-GNN achieves $21\times$-$71\times$ speedup compared with state-of-the-art systems after normalization.  }
HyScale-GNN achieves speedup for several reasons:
(1) resource utilization: HyScale-GNN utilizes both the processors and the accelerators to train GNN models collaboratively. 
In particular, HyScale-GNN utilizes both the CPU cores and the accelerators to compute; and utilizes both the CPU memory and device memory to read data concurrently.
Our DRM engine (Section \ref{sec:drm}) further ensures the tasks are efficiently mapped onto our platform.
On the other hand, PaGraph and $P^3$ do not take advantage of the processors on the platform.
(2) communication overhead: 
as mentioned in Section \ref{sec:perf_ev}, FPGA-based solutions can efficiently reduce the external memory access overhead compared with GPU-based solutions.
In addition, PaGraph only caches a portion of the vertex features in the device memory, and needs to fetch data from the CPU memory if it encounters a cache miss; thus, the PCIe communication overhead becomes large when training on large-scale graphs like ogbn-papers100M since cache miss occurs frequently.
$P^3$ incurs inter-node data communication since the graph is partitioned and distributed on each node, which causes extra communication overhead compared with HyScale-GNN.
Compared with DistDGLv2, which runs on eight nodes with a total of 64 GPUs, HyScale-GNN is able to achieve $0.45\times$ of its performance using only 4 FPGAs on a single node machine.
DistDGLv2 utilizes both the processor and the accelerator to train GNN models collaboratively. 
However, DistDGLv2 adopts a static task mapping, which can be inefficient.
In addition, DistDGLv2 partitions the input graph and distributes the partitions to each node, which incurs inter-node communication overhead like $P^3$.

\begin{table}[t]
\renewcommand{\arraystretch}{1.1}
\caption{Epoch time (sec) Comparison with State-of-the-art}
\label{tab:sota_perf}
\begin{tabular}{cccccc}
\toprule
 \multirow{2}{*}{} & \multicolumn{2}{l}{ogbn-products} & \multicolumn{2}{l}{ogbn-papers100M} & \multirow{2}{*}{\begin{tabular}[c]{@{}c@{}}Geo. mean\\ speedup\end{tabular}} \\ \cline{2-5}
                  & GCN           & SAGE         & GCN            & SAGE          &                                                                              \\ \midrule \midrule
PaGraph           & 1.18          & 0.25              & 4.00           & 1.18               & 1x                                                                           \\ 
This Work         & 0.27          & 0.49              & 0.58           & 1.91               & 1.76x                                                                        \\ \midrule
$P^3$                & 1.11          & 1.23              & 2.61           & 3.11               & 1x                                                                           \\ 
This Work         & 0.27          & 0.28              & 0.57           & 0.59               & 4.57x                                                                        \\ \midrule
DistDGL\_v2       & -             & 0.30              & -              & 4.16               & 1x                                                                           \\ 
This Work         & -             & 1.69              & -              & 3.67               & 0.45x                                                                                                                                                                                                                                                                            \\ \bottomrule
\end{tabular}
\end{table}

\begin{table}[ht]
\renewcommand{\arraystretch}{1.1}
\caption{{Normalized Epoch Time (Sec$\times$TFLOPS) Comparison with State-of-the-art}}
\label{tab:norm_perf}
\begin{tabular}{cccccc}
\toprule
 \multirow{2}{*}{} & \multicolumn{2}{l}{ogbn-products} & \multicolumn{2}{l}{ogbn-papers100M} & \multirow{2}{*}{\begin{tabular}[c]{@{}c@{}}Geo. mean\\ speedup\end{tabular}} \\ \cline{2-5}
                  & GCN           & SAGE         & GCN            & SAGE          &                                                                              \\ \midrule \midrule
PaGraph           & 135.1          & 28.63              & 458.0           & 135.1               & 1x                                                                           \\ 
This Work         & 2.59          & 4.70              & 5.55          & 18.34              & 21x                                                                        \\ \midrule
$P^3$                & 165.1          & 183.0              & 388.4           & 462.8               & 1x                                                                           \\ 
This Work         & 2.59          & 2.69              & 5.47           & 5.66             & 71x                                                                        \\ \midrule
DistDGL\_v2       & -             & 163.2            & -              & 2263               & 1x                                                                           \\ 
This Work         & -             & 16.20              & -              & 35.23               & 25x                                                                                                                                           \\ \bottomrule
\end{tabular}
\end{table}

\begin{figure}[t]
    \centering
    \includegraphics[width=8cm]{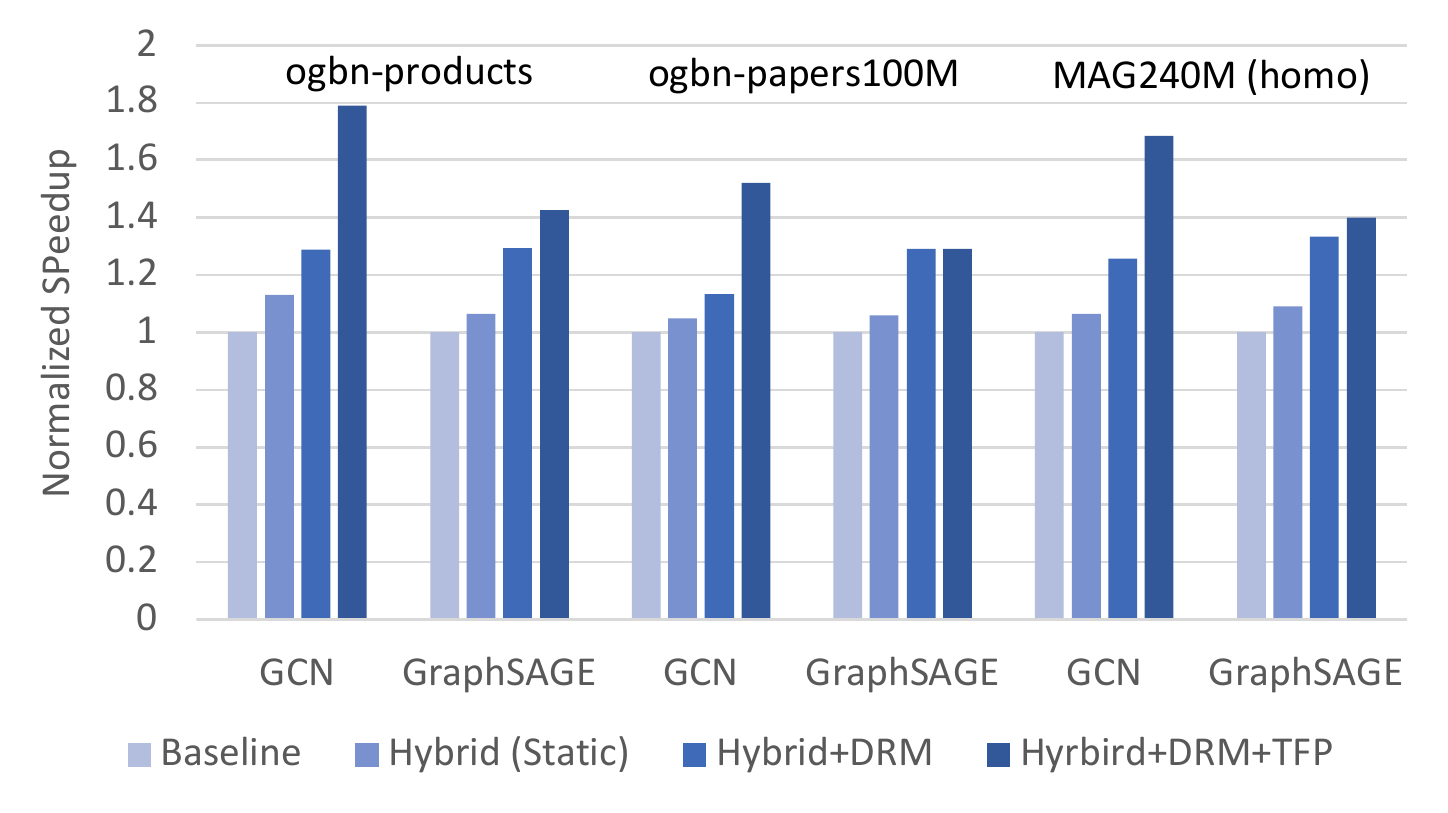}
    \caption{Impact of optimizations}
   
     \label{fig:opt}
\end{figure} 

\subsection{Ablation Study}\label{sec:ablation}
In this section, we evaluate the effectiveness of the optimizations applied in HyScale-GNN.
We show the evaluation on a CPU-FPGA heterogeneous architecture in Figure \ref{fig:opt};
evaluation on the CPU-GPU heterogeneous architecture also shows similar results.
We start from a baseline design, which adopts a traditional task mapping that offloads most of the tasks (except tasks like sampling, synchronization, etc.) to the FPGA.
Then, we apply hybrid CPU-FPGA training with a static task mapping; this leads to up to $1.13\times$ speedup.
The system achieves up to $1.33\times$ speedup after applying the DRM optimization (Section \ref{sec:drm}).
With the TFP (Section \ref{sec:pipe}) optimization applied, HyScale-GNN achieves up to $1.79\times$ speedup.
This is because the data loading stage is often a bottleneck in GNN training;
if the training is dominated by the GNN propagation stage (e.g., GraphSAGE model on the ogbn-papers100M in Figure \ref{fig:opt}), then the TFP optimization does not provide speedup.

\section{Related Work}
Several works have been proposed \cite{graphact,hp-gnn,rubik} to accelerate GNN training on a single node.
However, these works focus on using a single accelerator to perform GNN training and do not support training with multiple accelerators.
In addition, works like GraphACT \cite{graphact} and HP-GNN \cite{hp-gnn} stores the input graph in the device memory, and thus cannot support large-scale graphs \cite{hu2021ogblsc} that exceed the device memory size.
Recently, several works \cite{p3,distdgl,dgcl} have been proposed to train GNN on a multi-node platform.
However, these works require graph partitioning, which leads to issues like workload imbalance, and high inter-node communication overhead.
In addition, graph partitioning may affect the convergence rate and model accuracy \cite{distdgl}.
In this work, we show that it is feasible to train large-scale GNNs on a single node and achieve high training throughput.

\section{Conclusion}
In this work, we proposed HyScale-GNN, a hybrid training system that is optimized for training GNN models on large-scale graphs.
We proposed several optimizations to reduce the communication overhead and perform efficient task mapping.
Our system achieved up to $12.6\times$ speedup compared with a multi-GPU baseline.
In addition, using only four FPGAs on a single node, HyScale-GNN is able to achieve $1.76\times - 4.57\times$ speedup compared with state-of-the-art training systems that employ 8 to 16 GPUs.

We also observed some limitations of HyScale-GNN.
First, HyScale-GNN did not provide an effective solution if the performance is bottlenecked by the Data Transfer stage (i.e., limited by PCIe bandwidth).
In this case, the DRM engine would reduce the workload assigned to the accelerator, which limits the achievable speedup and scalability of the system.
Second, HyScale-GNN could not be directly extended to a distributed platform with multiple nodes.
If an input graph is partitioned and distributed to each node like in DistDGL \cite{distdgl}, inter-node communication and synchronization are needed.
However, our protocol defines how the processor and the accelerator should interact on a single node.
It does not support inter-node communication.
In the future, we plan to exploit techniques like data quantization to relieve the stress on the PCIe bandwidth, and define a more general protocol for training GNN models on distributed and heterogeneous architectures.

\ifCLASSOPTIONcompsoc
  \section*{Acknowledgments}
\else
  \section*{Acknowledgment}
\fi

{This work has been supported by the U.S. National Science Foundation (NSF) under grants SaTC-2104264 and OAC-2209563, and the DEVCOM Army Research Lab (ARL) under grant W911NF2220159.}



\end{document}